\documentclass[aps,pre,twocolumn, superscriptaddress]{revtex4}

\usepackage{graphicx}

\begin{document}

\title{Wavefunction correction scheme for non fixed-node diffusion Monte Carlo}


\author{Naz{\i}m Dugan}
\email{ndugan@metu.edu.tr}
\affiliation{Department of Physics, Middle East Technical University, 06531 Ankara, Turkey}

\author{\.{I}nan\c{c} Kan{\i}k}
\email{inanckanik@gmail.com}
\noaffiliation

\author{\c{S}akir Erko\c{c}}
\email{erkoc@metu.edu.tr}
\affiliation{Department of Physics, Middle East Technical University, 06531 Ankara, Turkey}

\date{\today}

\begin{abstract}
Wavefunction correction scheme, which was developed as a variance reduction tool for the pure and fixed-node diffusion Monte Carlo (DMC) computations by Anderson and Freihaut, is applied to the DMC computations of fermions without using the fixed-node constraint. This technique is found to be suitable for the non fixed-node calculations because of the significant decreases observed in the computation times for calculating results with certain statistical error values in the benchmark computations. 

\end{abstract}



\maketitle

\section{Introduction}
Antisymmetry condition of the wavefunction on identical particle exchanges complicates the electronic structure calculations of the fermionic many-body systems. The projector quantum Monte Carlo (QMC) methods \cite{Aspuru.Lester,Foulkes.Mitas}, facilitating imaginary time evolution of an initial quantum state, are accurate tools for such fermionic calculations. However, the evolution in imaginary time tends to the symmetric bosonic ground state instead of the antisymmetric fermionic ground state, resulting in the famous {\it fermion sign problem} \cite{Troyer}. Most of the attempts for an exact imposition of the antisymmetry condition in the projector methods facilitate plus and minus signed {\it walkers} diffusing and canceling each other whenever encounters occur \cite{Ceperley}. Usual population control mechanism in these methods can only stabilize the difference between the plus and minus signed populations and therefore cancellations of opposite signed walkers are essential for controlling the total population. Population control problem arises for larger systems since the plus and minus signed walker encounter rate is very low in the higher dimensional configuration spaces. The correlated dynamics of plus and minus signed walkers and the antisymmetric guiding functions used in the recently developed fermion Monte Carlo method \cite{Kalos} increase the cancellation rate to some extent. However, it was shown that these developments were not enough for the resolution of the sign problem \cite{Assaraf}. Therefore, applications of the exact methods are currently limited to a small number of fermions.

Wavefunction correction scheme was developed for the projector QMC computations as a variance reduction tool \cite{Anderson1}. The difference between the true ground state wavefunction and a trial wavefunction is sampled in this technique instead of the ground state wavefunction itself. This technique is used as an efficiency improvement in the projector QMC computations of bosons \cite{Anderson1, Anderson3, Anderson4} as well as fixed-node projector QMC computations of fermions \cite{Anderson4}.

In the current study, wavefunction correction scheme is applied to plus-minus cancellation facilitating computations of the fermionic systems without using the fixed-node approximation. Efficiency improvements due to the wavefunction correction scheme in such QMC calculations are investigated on some simple benchmark systems. Calculations are carried out using the diffusion Monte Carlo (DMC) \cite{Aspuru.Lester,Foulkes.Mitas,Kosztin} method but the correction scheme is applicable to other projector QMC methods as well.

\section{Method of Computation}

DMC, an highly accurate QMC method, relies on the fact that the form of the imaginary time Schr\"{o}dinger equation is a diffusion equation with a source term: 

\begin{equation}
	\partial_\tau \Psi({\bf x},\tau) = \frac{1}{2} 
			\nabla^2 \Psi({\bf x},\tau) - [V({\bf x})-E_R] \Psi({\bf x},\tau)~,
\end{equation}	

\noindent where ${\bf x}$ is the position vector in the configuration space of the physical system. The potential energy $V({\bf x})$ defines the interactions between the particles and with external sources. DMC treats the wavefunction $\Psi({\bf x},\tau)$ as a density distribution of some number of hypothetical particles, also called as walkers, diffusing in the $D \times N$ dimensional configuration space, $D$ being the number of space dimensions and $N$ being the number of identical particles. These walkers are subjected to a branching process according to the source term of the diffusion equation which is the term including the potential energy $V({\bf x})$ in the Schr\"{o}dinger equation. Population control is established by controlling the rate of the branching process via adjustments of the reference energy $E_R$ which is an overall energy shift. The evolution in imaginary time $\tau$ projects out the ground state component of an arbitrary initial wavefunction $\Psi({\bf x},0)$ in the long $\tau$ limit \cite{Kosztin}. The DMC method uses short time propagator and thus the calculated expectation value result has a time step error which can be made insignificant by a time step extrapolation.

When the fixed-node constraint is not enforced the sign problem manifests itself as the problem of imposing the antisymmetry condition. Minus signed walkers arise in such non fixed-node calculations, breaking down the population control mechanism for large systems even with the cancellations of opposite signed walkers.

In the correction scheme, DMC method is modified to make correction on a known trial wavefunction \cite{Anderson1, Anderson3, Anderson4}. Wavefunction is divided into two parts as the trial wavefunction $\Phi_T({\bf x})$ and the remaining unknown part $\Phi({\bf x},\tau)$ which is sampled through the DMC calculation. It is helpful to substitute  $\Phi({\bf x},\tau)+\Phi_T({\bf x})$ for the wavefunction $\Psi({\bf x},\tau)$ in the imaginary time Schr\"{o}dinger equation for comprehending the effect of the correction scheme:

\begin{eqnarray}
\begin{array}{ll}
\partial_\tau \Phi({\bf x},\tau) &= \frac{1}{2} 
			\nabla^2 [\Phi({\bf x},\tau) + \Phi_T({\bf x})] \\
				  & \\ 	
				  & - [V({\bf x})-E_R] [\Phi({\bf x},\tau) + \Phi_T({\bf x})]~,
\end{array}
\end{eqnarray}					     	        

\noindent which can be simplified using the definition of the local energy $E_L({\bf x}) = \hat{H} \Phi_T({\bf x})/\Phi_T({\bf x})$ as follows:

\begin{eqnarray}
\begin{array}{ll}
\partial_\tau \Phi({\bf x},\tau) &= \frac{1}{2} 
			\nabla^2 \Phi({\bf x},\tau) - [V({\bf x})-E_R] \Phi({\bf x},\tau)\\
				  & \\
				  & -[E_L({\bf x})-E_R]~\Phi_T({\bf x}). 	
\end{array}
\end{eqnarray}

\noindent Last term on the rhs of the above equation is the additional term related to the correction scheme whose sole effect may be simulated by some number of vacuum branchings carried out along the simulation region with the branching factors linearly proportional to the extra term $[E_L({\bf x})-E_R]~\Phi_T({\bf x})$. A plus or minus signed walker depending on the sign of this term may be added to the walker population as a result of a single vacuum branching. In the current computations, some number of points in the configuration space are generated  using the Metropolis algorithm according to the distribution $\Phi_T({\bf x})$ which is positive definite in the simulation region described below and the branching factors are calculated using the factors $[E_L({\bf x})-E_R]$.

Antisymmetry condition is imposed on the wavefunction by using the concept of the permutation cell which is the repeating unit cell of the wavefunction for identical fermions and bosons \cite{Anderson2,Luczac}. Computation is carried out in a single permutation cell which is taken as a positive valued nodal region of the trial wavefunction $\Phi_T({\bf x})$ (Trial wavefunctions are chosen to have nodal regions having permutation cell property. Density functional theory results also have this property \cite{Ceperley2}). All the walkers of the plus and minus sign are initially generated in the chosen permutation cell and the outgoing walkers are permuted back to the simulation region. A sign change is applied if an odd number of particle permutations are required to take the walker back inside the permutation cell.

The normalization of the wavefunction is satisfied by keeping the plus and minus signed walker amounts equal to each other via step by step adjustments of the DMC reference energy which affects the rates of the walker and vacuum branchings. The trial wavefunction used for the correction is normalized separately to an optimum value. The ratio of the trial wavefunction normalization ($\int \Phi_T({\bf x})d\Omega $ :integral is over the simulation region where $ \Phi_T({\bf x})$ is positive) to the number of walkers from each sign is an important parameter ($r_{\mbox{\scriptsize{n}}}$) of the correction scheme calculations. A larger value of this ratio increases the efficiency of the correction scheme calculation since the contribution of the walker distribution and the variance of the computed result decreases in such a case. However, if the ratio is adjusted to have a very high value the effect of the walker distribution gets insignificant and the wavefunction is not corrected. This mentioned ratio and consequently the efficiency of the method can be increased as the trial wavefunction gets closer to the true fermionic wavefunction. 

Population control is established by the cancellations of opposite signed walkers which encounter during the correlated random walk process in which the Gaussian random walk vectors of the paired walkers are symmetric with respect to the perpendicular bisector of the line connecting the two walkers  \cite{Kalos}. Cancellation process is temporarily stopped  when the number of walkers decreases below a certain threshold value for the small dimensional calculations (up to four dimensional configuration spaces) since the cancellation rate is very high for them.   
    
Computation of the ground state energy expectation value should be modified accordingly. Necessary modifications are derived by integrating both sides of the eigenvalue equation over the simulation region volume. The final energy expression is:

\begin{widetext}
\begin{equation}
E = \frac{-\frac{1}{2}\int_{\partial \Omega} \nabla 
						 \Phi({\bf x},\tau).dS
				    + \int_{\Omega} V({\bf x})\Phi({\bf x},\tau)d\Omega + \int_{\Omega} E_L({\bf x}) \Phi_T({\bf x})d\Omega}{\int_{\Omega}[\Phi({\bf x},\tau)+\Phi_T({\bf x})]d\Omega}~,
\end{equation} 
\end{widetext}

\noindent where divergence theorem is used to aquire the first term of the numerator which is about the walker flow at the boundaries of the simulation region \cite{Luczac}. Second term of the numerator is the sum of walker potential energies in the energy calculation of the usual DMC. Third integral of the numerator is about the trial wavefunction being corrected throughout the computation and it can be calculated in the beginning of the simulation using Monte Carlo integration technique without respecting the normalization of the trial wavefunction. The value of the integral $\int \Phi_T({\bf x})d\Omega $ of the denominator is given as a parameter of the method (determines the parameter $r_{\mbox{\scriptsize{n}}}$ together with the number of walkers from each sign) and the mentioned Monte Carlo integration is multiplied by this given value of the $\Phi_T({\bf x})$ integral for the normalization issue. The remaining first integral of the denominator is the difference between the number of plus and minus signed walkers. Ground state energy is calculated considering these separate terms in each time step and it is time averaged after some thermalization steps. The time step errors are ignored in the following benchmark calculations since they are very small compared to the statistical error bars.

\section{Benchmark Computations}
\subsection{Harmonic fermions}
Harmonic fermions are preferred in the benchmark calculations for their property of being exactly solvable. Analytical solutions are disturbed slightly to prepare trial wavefunctions suitable for testing the current method. Two fermion systems are studied which have Hamiltonian functions in the following form:

\begin{equation}
H = -\frac{1}{2} (\nabla^2_1 + \nabla^2_2) + \frac{1}{2} \omega^2~({\bf r}_1^2+{\bf r}_2^2),
\end{equation} 

\noindent where the vectors ${\bf r}_1$ and ${\bf r}_2$ denote individual particle coordinates and $\omega^2$ is a constant whose numerical value is taken as $0.03$ in the current calculations. The trial wavefunction is chosen as:

\begin{equation} 
\Psi_{T} = e^{\varepsilon_1 \frac{\omega}{2}({\bf r}_1^2+{\bf r}_2^2)} (x_2 + \varepsilon_2~ y_2^2 - x_1 - \varepsilon_2~y_1^2),
\end{equation}

\noindent where $x$, $y$ are the particle coordinate components in two separate space dimensions and $\varepsilon_1$, $\varepsilon_2$ are free parameters. This trial wavefunction gives the true fermionic ground state in $\varepsilon_1 \rightarrow 1$, $\varepsilon_2 \rightarrow 0$ limit for arbitrary number of space dimensions. A non zero value of the parameter $\varepsilon_2$ distorts the nodal surface of the trial wavefunction for space dimensions larger than one. However, the permutation cell property of the nodal region is preserved which is taken as the simulation region where outgoing walkers are taken inside as described in the previous section. The DMC time step is chosen as 0.003 dimensionless time units and data is collected for 80000 time steps after the thermalization steps. Number of points for the vacuum branching process is chosen as 500 and kept constant during the computations. The ratio parameter $r_{\mbox{\scriptsize{n}}}$ values are set to the highest values allowing the complete correction for each case since the efficiency of the correction scheme increases as the $r_{\mbox{\scriptsize{n}}}$ increases. These values are easily determined since a deviation in the calculated energy expectation value starts to occur beyond a certain point as the mentioned ratio increases. Calculated energy expectation value matches with the true value below this point which is identified as the optimum value of the parameter $r_{\mbox{\scriptsize{n}}}$.

Computations are also carried out without using the correction scheme for a comparison of the computational efforts of the two cases. DMC without any corrected trial wavefunctions is used for these comparison calculations. Same permutation cells that used in the correction scheme computations are used where outgoing walkers are treated in the same way. Correlated walk of opposite signed walkers with the cancellation process is also facilitated in the comparison case computations. Computation times of the two cases for calculating the results with certain statistical error values are compared.  

Computation results for the harmonic fermion systems up to four space dimensions are given in TABLE \ref{harmonic.results} for the correction scheme computations. Energy expectation values for the used trial wavefunctions ($E_{\mbox{\scriptsize{T}}}$), calculated by Monte Carlo integration technique, are also given. Computed energies are very close to the true values ($E_{\mbox{\scriptsize{GS}}}$) for the all cases. Efficiency improvements can be seen from the ratios of the comparison case computation times to the correction scheme computation times.  Significant decreases in the computation times are observed for the all studied space dimensions when the correction scheme is used. 

\begin{table}[h]
 \caption{Correction scheme computation results for two harmonic fermions. $d$: space dimension, $\varepsilon_1$,$\varepsilon_2$: disturbance parameter values, $E_{\mbox{\scriptsize{c}}}$: calculated energy expectation value using the correction scheme, $E_{\mbox{\scriptsize{GS}}}$: true value of the fermionic ground state energy, $E_{\mbox{\scriptsize{T}}}$: trial wavefunction energy (all energies are given in dimensionless units), $N_{\mbox{\scriptsize{w}}}$: stabilized number of walkers from each sign, $r_{\mbox{\scriptsize{n}}}$: ratio of the trial wavefunction normalization to the number of walkers from each sign, $r_{\mbox{\scriptsize{t}}}$: ratio of the comparison case computation time to the correction scheme computation time.}
  \label{harmonic.results}
  \begin{center}
  \begin{tabular}{lllllllll}
    \hline
    \hline
    $d$ & $\varepsilon_1$ & $\varepsilon_2$ & $E_{\mbox{\scriptsize{c}}}$ & $E_{\mbox{\scriptsize{GS}}}$ & $E_{\mbox{\scriptsize{T}}}$ & $N_{\mbox{\scriptsize{w}}}$ & $r_{\mbox{\scriptsize{n}}}$ & $r_{\mbox{\scriptsize{t}}}$ \\
    \hline
    1 & 0.964 & 0.00 & 0.34644(31) & 0.34641 & 0.34677 & 195 & 14.4 & 2.7\\
    2 & 1.000 & 0.05 & 0.51978(62) & 0.51962 & 0.52205 & 495 & 5.7 & 5.5\\
    3 & 1.000 & 0.05 & 0.6932(11) & 0.69282 & 0.70143 & 520 & 4.6 & 8.8\\
	 4 & 1.000 & 0.05 & 0.8660(10) & 0.86603 & 0.87473 & 1556 & 1.9 & 4.8\\
    \hline
    \hline
  \end{tabular}
  \end{center}
\end{table}

\begin{figure}[h!]
 \includegraphics[angle=-90,scale=0.3]{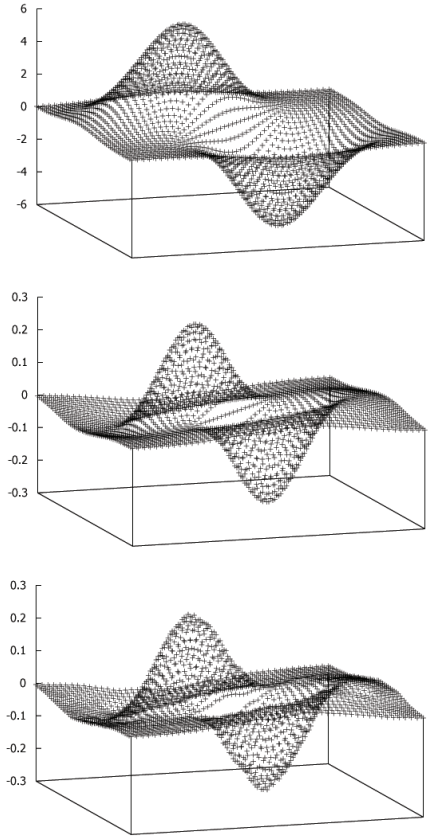}
 \caption{Wavefunction images for correction scheme DMC computation of the two   harmonic fermions in 1D. {\bf TOP:} Trial wavefunction. {\bf MIDDLE:} Difference between the trial wavefunction and the true fermionic ground state. {\bf BOTTOM:} Average walker distribution during the DMC computation.}
 \label{1D}
 \end{figure}

Images for the trial wavefunction (top image) and its difference from the true fermionic wavefunction (middle image) is given in the FIG. \ref{1D} for the 1D computation whose configuration space is two dimensional. Average walker distribution during the correction scheme DMC computation (bottom image) is also given. Walker distribution is calculated in a single permutation cell and it is reflected to the other cell with a sign inversion in order to generate a plot for the all configuration space. Minus signed walkers give negative weights when the average is calculated. Walker distribution fits well with the difference function as expected when the correction scheme is used.

\subsection{Helium atom lowest triplet state: 1s2s $^3$S}

As a more realistic example, Helium atom lowest triplet state energy eigenvalue is computed using the current method. Non-relativistic Hamiltonian operator with the coulombic interparticle interaction is used. Trial wavefunction to be corrected is a Slater determinant taken from the work of Emmanouilidou et al. \cite{Emmanouilidou}:

\begin{equation}
\Psi_T=e^{-(2 r_1 + \alpha r_2)} (1-\alpha~r_2)
		-e^{-(2 r_2 + \alpha r_1)} (1-\alpha~r_1)~, 
\end{equation}

\noindent where $r_1$ and $r_2$ are the electron nucleus distances for the two electrons of the system and the numerical value of the parameter $\alpha$ is chosen as 0.65. This trial function does not satisfy the true nodal surface of the 1s2s $^3$S state of the Helium atom which is well known as $r_1=r_2$ surface. Plots of the trial wavefunction at various cross sections show that the nodal surface of the trial wavefunction extends out of the true nodal surface. 

DMC time step is chosen as 0.0005 atomic time units and the data is collected for 80000 time steps. Initial number of plus and minus signed walkers and the number of points for the vacuum branchings are set to 500. The normalization of the trial wavefunction is set to 2600 and the number of walkers of each sign is stabilized around 520, giving the $r_{\mbox{\scriptsize{n}}}$ value as 5.0. 

The non-relativistic true energy expectation value for this state is -2.1752 Hartrees \cite{Bailey} and the energy expectation value of the used trial wavefunction is -2.1548 Hartrees. Current correction scheme calculation gives the result as -2.1767(63) Hartrees for which the true value is in the statistical error interval. Computation time of the correction scheme calculation is 4.2 times shorter than the computation time to achieve the same precision with the comparison case calculation without any trial wavefunctions. Fixed-node DMC calculation using the nodes of the used trial wavefunction gives the energy value as -2.1626(8) Hartrees confirming the deviation of the trial wavefunction nodal surface from the true nodal surface $r_1=r_2$. 
\\
\\
\section{Discussion}

Application of the wavefunction correction scheme to the non fixed-node DMC increases the efficiency of the method significantly. Benchmark computations on the harmonic fermions and the helium atom show that the computation time for calculating the result within a certain statistical error value decreases several times when the correction scheme is used. Improvement of the efficiency depends on the ratio of the trial wavefunction normalization to the number of walkers and this parameter of the method can be increased when the trial wavefunction gets closer to the true fermionic wavefunction. Therefore, quality of the trial wavefunction is an important factor affecting the efficiency improvement observed in the correction scheme calculations. The mentioned ratio parameter value should be set to the highest value allowing a full correction which can be guessed considering the quality of the trial wavefunction. This parameter value can be optimized using the fact that a deviation in the calculated expectation value starts to occur beyond the optimum value of the parameter. 

The nodal surfaces of the used trial wavefunctions deviate from the true surfaces for the all cases except the harmonic fermions calculation in 1D for which a permutation cell preserving node distortion is not possible. The correction scheme calculations yield the true energy expectation value despite the wrong nodal surfaces which proves the applicability of the wavefunction correction scheme for the non fixed-node QMC calculations.  

Fixed-node DMC is applicable to large systems since it eliminates the minus signed walkers by constraining the computation in a nodal region. This advantage of the fixed-node approximation disappears with the usage of the correction scheme because of the arising minus signed walkers as a result of the vacuum branchings. However, the wavefunction correction technique is suitable for non fixed-node DMC computations since the minus signed walkers are already needed for plus-minus cancellation methods. Correction scheme improves the large scale applicability for such calculations as opposed to the case for the fixed-node calculations.

Benchmark computations are carried out without using the importance sampling transformation in order to better observe the sole effect of the wavefunction correction technique. Importance sampling may be facilitated in the correction scheme calculations as described in the references \cite{Anderson3, Anderson4}. However, the guiding function should allow the walkers' passage from the boundaries of the chosen permutation cell for the application of the current boundary conditions when the fixed-node constraint is relaxed. A slightly modified form of the corrected trial wavefunction which does not vanish on the nodal surface of the original function may be used as the guiding function.

A generally applicable method beyond the fixed-node approximation is very desirable for high accuracy electronic structure calculations of relatively larger systems. Current non fixed-node QMC methods have exponential scaling computation cost with increasing  number of fermions and therefore not applicable to large systems. The fermion sign problem may not have a polynomial time solution with the classical computation techniques. However, fixed-node constraint may be relaxed by some sort of other approximate manners and the application of the non fixed-node computations may be widened by the usage of improvements like the wavefunction correction technique used in the current study.  

\acknowledgments{Authors thank J. B. Anderson and R. Assaraf for valuable discussions about the subject.}

\bibliography{csdmc}

\end{document}